\documentclass[chicago]{emulateapj}
\usepackage{graphicx,amsmath,natbib,bm}
\usepackage[none]{hyphenat}
\usepackage{amsfonts}
\usepackage[top=1.3in, bottom=-0.3in, left=0.9in, right=0.35in]{geometry}
\usepackage{times}
\shortauthors{Y. Hezaveh}
\begin{document}

\title{Prospects for measuring the mass of Black Holes at high redshifts with\\
 resolved kinematics   using Gravitational Lensing}
\author{Yashar D. Hezaveh}  
\affil{Kavli Institute for Particle Astrophysics and Cosmology, Stanford University, Stanford, CA, USA}

\begin{abstract}  
\noindent
Application of the most robust method of measuring black hole masses, spatially resolved kinematics of gas and stars, is presently limited to nearby galaxies.
The Atacama Large Millimeter/sub-millimeter Array (ALMA) and thirty meter class telescopes (the Thirty Meter Telescope, the Giant Magellan Telescope, and the European  Extremely Large Telescope) with milli-arcsecond resolution are expected to extend such measurements to larger distances.
Here, we study the possibility of exploiting the angular magnification provided by strong gravitational lensing to measure black hole masses at high redshifts ($z\sim 1-6$), using resolved gas kinematics with these instruments. 
We show that in $\sim15\%$ and $\sim20\%$ of strongly lensed galaxies, the inner $25$ and $50$ pc could be resolved, allowing the mass of $\gtrsim 10^8 M_{\odot}$ black holes to be dynamically measured with ALMA, if moderately bright molecular gas is present at these small radii.
 Given the large number of strong lenses discovered in current millimeter surveys and future optical surveys, this fraction could constitute a statistically significant population for studying the evolution of the $M-\sigma$ relation at high redshifts.\\
\end{abstract}


\section{introduction}
One of the most remarkable discoveries of modern astrophysics is the finding that almost every massive galaxy harbors a super massive black hole (SMBH) at its center and that the mass of the central black hole (BH) is proportional to the mass of its host galaxy \citep{Kormendy:95,Magorrian:98,Ferrarese:00,Gebhardt:00,Tremaine:02,Kormendy:13}. This tight correlation suggests an intriguing physical connection between galaxy and BH formation throughout the history of the universe \citep{Silk:98,Fabian:99,King:03}. Uncovering the nature of the physical processes which induce this correlation remains one of the most exciting goals of modern astrophysics.

A crucial step toward identifying these mechanisms is understanding the evolution of this correlation at high redshifts during the epochs when today's galaxies were actively forming their stellar populations.
In the local universe, the most robust tool for measuring BH masses has been spatially resolved kinematics of gas and stars \citep[e.g.][]{Gultekin:09}. 
At high redshifts, unfortunately, it has been impossible to employ this method since an extremely high angular resolution is needed to resolve the kinematics of the inner few parsecs of galaxies.
In the absence of such a possibility, the only available method has been the ``virial technique'' \citep{Ho:99,Wandel:99,Kaspi:00}, which uses the size and line width of broad-line regions in 
active galactic nuclei (AGNs), calibrated using reverberation mapping techniques \citep{Blandford:82,Peterson:93}.
This method, however, is only applicable to bright AGNs with broad line emission
and cannot be used to measure BH masses in typical galaxies.

 In this work, we study the possibility of measuring BH masses at high redshifts ($z\gtrsim 2$) using spatially resolved gas kinematics, taking advantage of the angular magnification provided by gravitational lensing. Gravitational lensing magnifies the observed angular 
sizes of background galaxies, in effect acting as a natural telescope \citep[e.g.][]{Yun:97}. 
Recently, \citet{Davis:13} reported a BH mass measurement ($M_{\mathrm{BH}}= 4.5^{+4.2}_{-3.0}\times 10^8 M_{\odot}$) using resolved kinematics of CO in NGC4526 obtained with CARMA. 
Atacama Large Millimeter/sub-millimeter Array (ALMA) and thirty meter class optical telescopes are expected to achieve angular resolutions of the order of milli-arcsecond and will be capable of spatially resolved spectroscopy, extending SMBH measurements to significantly larger distances \citep[e.g.][]{Davis:2014, Do:14}.
An angular resolution of $\sim 20$ mas translates into a physical resolution of $\sim 170$ pc at $z=2$, only a small factor ($\sim 3$) larger than the sphere of influence of a $5\times 10^8 M_{\odot}$ BH. A modest lensing magnification can allow these scales to be resolved and the masses of central BHs at high redshifts to be securely measured.
Motivated by the measurement of \citet{Davis:13}, the imminent completion of ALMA, and the approaching arrival of Thirty Meter Telescope (TMT), Giant Magellan Telescope (GMT), and European Extremely Large Telescope (E-ELT), we examine the possibility of measuring BH masses in lensed galaxies at $z\sim 1-6$ using molecular gas and quantify the probability of such measurements.

Thanks to wide area millimeter surveys (e.g., the South Pole Telescope and \emph{Herschel} space telescope), large populations of submillimeter-bright, high redshift, strongly lensed galaxies are now being identified and confirmed \citep{hezaveh:13b,vieira:13,bussmann:13}. It is expected that hundreds of strong lenses will be discovered in these surveys \citep{hezaveh:11}.
It is also expected that the new generation of optical and radio surveys will discover thousands of galaxy-galaxy strong lenses \citep[e.g. $\gtrsim1500$ with the Large Synoptic Survey Telescope and $\sim10^4$ with the Square Kilometer Array; ][McKean et al. 2014 in preparation]{LSST:09}.
Even a small, high-magnification fraction of such a large population could constitute a statistically significant sample of measured BH masses.

In Section \ref{sec:sim}, we describe the simulations we use to quantify ALMA's ability to detect the masses of lensed BHs.
In Section \ref{sec:results} we investigate the configurations which allow BH masses to be measured and estimate the probability of encountering such systems.
We discuss the results in Section \ref{sec:discuss} and finally conclude in Section \ref{sec:conclusion}.
 Throughout this work, we have assumed a flat $\Lambda$CDM cosmology with $h=0.71$ and $\Omega_M=0.26$.

\section{Simulations}
\label{sec:sim}

\begin{figure}
\begin{center}
\centering
\includegraphics[trim = 2 0 0 0 cm, clip, width=0.48\textwidth]{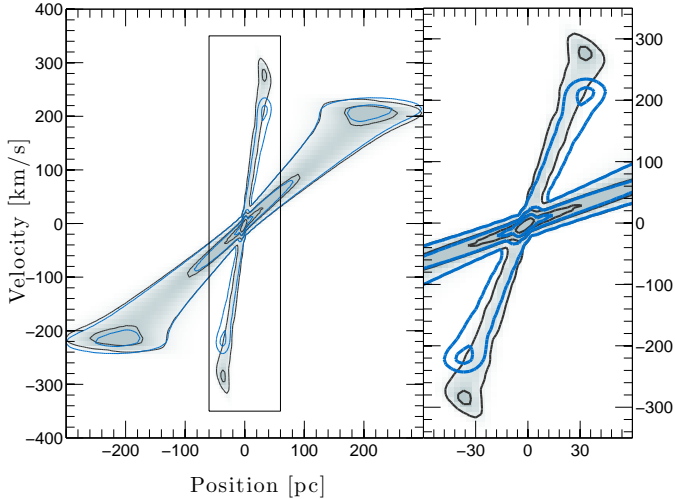}
\centering
\end{center}
\caption{Gas model in the source: plot of line-of-sight velocity of gas as a function of physical distance from the galaxy center. The model galaxy has two gas rings (similar to observations of Davis et al. 2013).
The black (shaded) curves show the mock data including a central black hole ($M_{\mathrm{BH}}=5\times 10^8$). The blue curves show the best fit model without a black hole. The Keplerian rise of the velocity in the inner ring (right panel) is used to measure the mass of the central black hole.
\label{f:f1}}
\end{figure}

We generate mock data cubes of gas emission in the source galaxy, lens the data cubes with a foreground halo, predict the ALMA visibilities, and use the visibilities to estimate the detection significance of the mass of a central BH for various parameters. 
The gas in the source galaxy is modeled as two circular annuli with radii of 50 pc and 200 pc, a Gaussian radial profile of thickness of 20 and 160 pc (FWHM), 
and an intrinsic velocity dispersion  of 10 $\mathrm{km\, s}^{-1}$ (FWHM) similar to observations of \citet{Davis:13}. 
The choice of two discrete rings is made to enforce resolving the inner 50 pc region for a BH mass measurement. We point out, however, that assuming a more general case (e.g., an exponential disk) does not change the results \citep[see Figure 2 of ][]{Davis:2014}.
The velocity integrated flux of the circumnuclear ring is set to 4 $\mathrm{mJy\, km\, s}^{-1}$. 
This value is calculated by scaling the flux of the circumnuclear CO gas in Arp 220 to $z=2$ \citep{Sakamoto:99}.
The rings are concentric and in circular orbits. The rotational velocity of each point in the rings is calculated based on the enclosed mass contributed from the galaxy mass model and the mass of the central BH. The BH is given a mass of $M_{\mathrm{BH}}=5\times 10^{8} M_{\odot}$.
We model the galaxy density with an isothermal profile ($\rho\propto r^{-2}$) with a central velocity dispersion of 200 $\mathrm{km\, s}^{-1}$. 
The line-of-sight velocities are calculated to generate a frequency-position data cube. Figure \ref{f:f1} shows a velocity-position diagram extracted from the data cube. The data comprises 100 frequency channels covering 800 $\mathrm{km\, s}^{-1}$.

Each layer of the data cube is lensed with the foreground halo to  generate a lensed cube. We use a singular isothermal ellipsoid (SIE) mass model for the lens galaxy \citep[$\rho\propto r^{-2}$,][]{Kormann:94}, placed at $z_l=0.5$, and place the source at $z_s=2.0$. The central velocity dispersion of the lens is set to 180 $\mathrm{km\, s}^{-1}$, which is typical for galaxy-galaxy lensing systems \citep[e.g.][]{hezaveh:13b}.

To predict the visibilities, we calculate the ALMA $uv$ coverage for a five hour long observation with the most extended antenna configuration (full array), using the $simobserve$ task of the Common Astronomy Software Applications package, which results in an angular resolution of $\sim$20 mas at an observing frequency of 240 GHz. 
The visibilities for each channel are calculated by computing the two-dimensional (2D) Fourier transform of the corresponding layer of the data cube and resampling the Fourier transform maps over the $uv$ coverage 
The noise is estimated using the ALMA sensitivity calculator for a channel width of 8 $\mathrm{km\, s}^{-1}$ at 240 GHz. 
We use finite differencing of visibilities to calculate the Fisher information and the parameter covariance matrix to calculate the significance of the BH mass measurement.

When the background source only consists of the two rings described above, the lens model parameters have large uncertainties and are mildly degenerate with the mass of the BH. 
A realistic source galaxy, however, will consists of many other components, extended over kiloparsec scales, which, if included in lens models, can strongly constrain the lens parameters.
We run five simulations in which the background source is embedded in an extended Gaussian clump with a radius of 1 kpc and a flux of 1 $\mathrm{Jy\, km\, s}^{-1}$ \citep[typical for SMGs; see][]{Bothwell:12}. We find that in these simulations, the lens parameters are extremely constrained and the effect of marginalization over them barely reduces the significance of BH mass measurements ($\lesssim 0.05\, \sigma$). This suggests that when the full structure of the source galaxies are included in models, the lens parameters do not significantly contribute to the BH mass uncertainties. Therefore, to reduce the computational cost of simulations, we do not marginalize over the lens parameters.

\section{Results}
\label{sec:results}

\subsection{BH Mass Measurements}

When the lensing angular magnification is large enough such that the inner ring is resolved, the measured velocities of gas at $r=50$ pc in combination with that of the outer ring (at $r=200$ pc) show a Keplerian rise of the rotational velocities, allowing us to simultaneously constrain the depth of the potential of the galaxy (central velocity dispersion in case of an isothermal profile, $\sigma_{v}$), and the mass of the central BH. 
 We use the Fisher information to compute the covariance matrix of the model parameters to estimate the detection significance of the mass of the central BH in a given mock observation. For a set of parameters, we simulate a mock observation following the procedure described in Section \ref{sec:sim}. 
The background source is parameterized by the position, radius, width, and flux of the rings,  the mass of the BH, and the depth of the potential of the host galaxy, $\sigma_{v}$. The lens parameters are fixed to their true values (see discussion in Section \ref{sec:sim}).

Resolving the inner ring requires a lensing magnification along the direction of the velocity field (perpendicular to the rotation axis). The lensing magnification is not isotropic: strong lensing magnifies along different dimensions with different factors, resulting in shear.
For a lens with an isothermal density profile, for example, in certain positions behind the lens the angular magnification along the $y$-axis is close to one (no magnification), while a large magnification along the $x$-axis exists. In that case, if the rotational axis of the ring is oriented horizontally, there will be no angular magnification along the velocity structure of the ring and consequently the ring is not resolved. If the rotational axis of the ring is oriented vertically, however, the large magnification along the velocity gradient of the source will allow a measurement of the velocity and radius of the ring.

\begin{figure}[h]
\begin{center}
\centering
\includegraphics[trim = 5 0 5 2 cm, clip, width=0.429\textwidth]{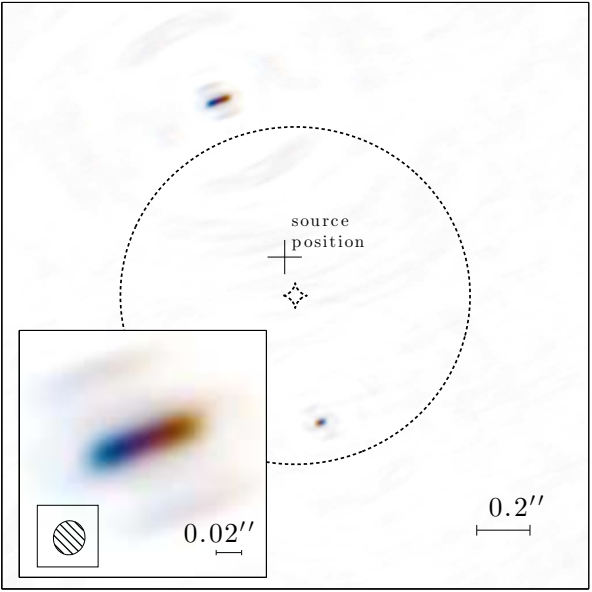}
\centering
\end{center}
\caption{ Example of simulated ALMA image (only the inner, $R=50$ pc ring shown), resulting in a $\sim3.7\sigma$ detection of the central BH mass.
The linear magnification of the upper image along the velocity gradient is $\sim 5$. ALMA beam is shown as the hatched area in the zoom panel.
\label{f:image}}
\end{figure}

\begin{figure}[h]
\begin{center}
\centering
\includegraphics[trim = 3 2 6 0 cm, clip , width=0.5\textwidth]{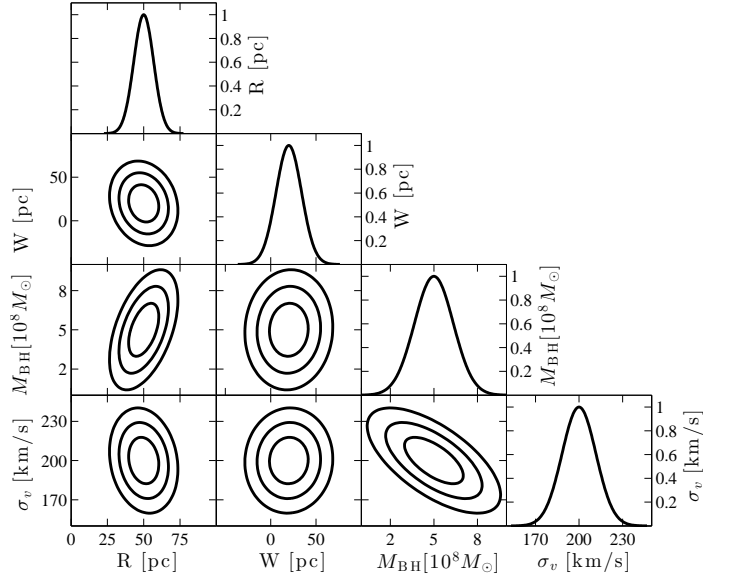}
\centering
\end{center}
\caption{Fisher forecast for the measurement of the radius ($R$) and width ($W$) of the ring, the mass of the BH ($M_{\mathrm{BH}}$), and the central velocity dispersion of the galaxy ($\sigma_v$) for a lensed galaxy at $z=2$, with ALMA after marginalizing over nuisance parameters.
The plot corresponds to the configuration illustrated in Figure \ref{f:image}.
\label{f:fisher}}
\end{figure}

We find that when the angular magnification of one of the lensed images along the velocity gradient of the source galaxy is large enough for the inner ring to be resolved, the mass of the  BH could be measured with high significance. Figure \ref{f:image} shows a simulated ALMA observation (dirty image) of a system in which the inner ring is resolved (only the inner ring is shown) with moderate angular magnification ($\mu_A\sim5$).  Figure \ref{f:fisher} shows the parameter covariance of this simulation with a marginalized BH mass measurement with a significance of $3.7 \sigma$.

\subsection{Probability of BH Mass Measurements}

Since measuring the BH mass requires specific conditions (a certain minimum magnification along the direction of the velocity field), it is possible that only a small number of lensed galaxies are suitable for such measurements.
In this section, we evaluate the probability of lensing configurations which allow the mass of central BHs to be measured.

\begin{figure*}
\begin{center}
\centering
\includegraphics[trim = 0 0 0 0 cm, width=0.6\textwidth]{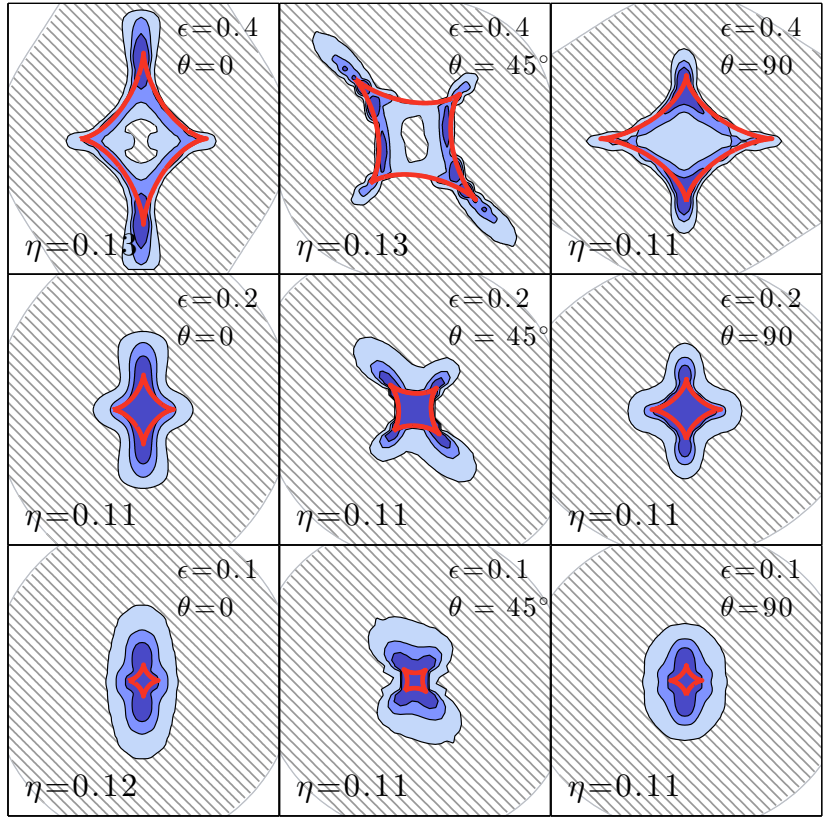}
\centering
\end{center}
\caption{Blue-shaded contours (light to dark) show the marginalized 3$\sigma$, 5$\sigma$, and 7$\sigma$ detection cross sections for a $5\times10^8$ $M_\odot$ black hole.  
The dashed areas shows the $\mu_f=3$ flux magnification cross sections. 
The panels correspond to different lens orientation angles (left to right, $\theta=0^{\circ}$, $45^{\circ}$, and $90^{\circ}$) and ellipticities (bottom to top, $\epsilon=0.1$, 0.2, and 0.4). The red curves show the diamond caustic of the lens.
\label{fig:cross}}
\vspace{0.5cm}
\end{figure*}

We run a set of simulations in which we place the source at various positions (a set of points on a Cartesian grid) behind the lens. At each point, using Fisher analysis, we evaluate the significance of the BH mass measurement. This allows us to measure a detection significance cross section for a lens.
We repeat the procedure for different mass parameters (ellipticity and orientation angle). 
The gray shaded contours in Figure \ref{fig:cross} show the 3$\sigma$, 5$\sigma$, and $7\sigma$ detection cross sections ($C_{3\sigma}, C_{5\sigma}, C_{7\sigma}$) for three lens ellipticities ($\epsilon$=0.1, 0.2, and 0.4) and orientation angles ($\theta$ = 0, 45, and $90^{\circ}$).  

The hatched areas in Figure \ref{fig:cross} show the $\mu_{f}\geqslant3$ flux magnification cross section, $C(\mu_f\geqslant3)$. 
This cross section is the area behind a lens inside which the lensed flux of the background galaxy (combined flux of all images) is boosted by a factor greater than $\mu_f$. 
The population of strongly lensed sources detected in millimeter surveys have been shown to have large flux magnifications, with almost all sources having $\mu_f\geqslant3$ \citep{hezaveh:12a,hezaveh:13b}.
 In other words, the background sources in the sample of strongly lensed galaxies from these surveys reside inside the $C(\mu_f\geqslant3)$ cross section (hatched contours).
 We define the yield factor, $\eta$, as the ratio of the area of the $3 \sigma$ BH detection, $C_{3\sigma}$, to the area of the lensing cross section $C(\mu_f\geqslant3)$ which indicates the fraction of strongly lensed sources with suitable configurations for central BH mass measurements.

\begin{figure}[h]
\begin{center}
\centering
\includegraphics[trim = 3 0 0 0 cm, clip, width=0.48\textwidth]{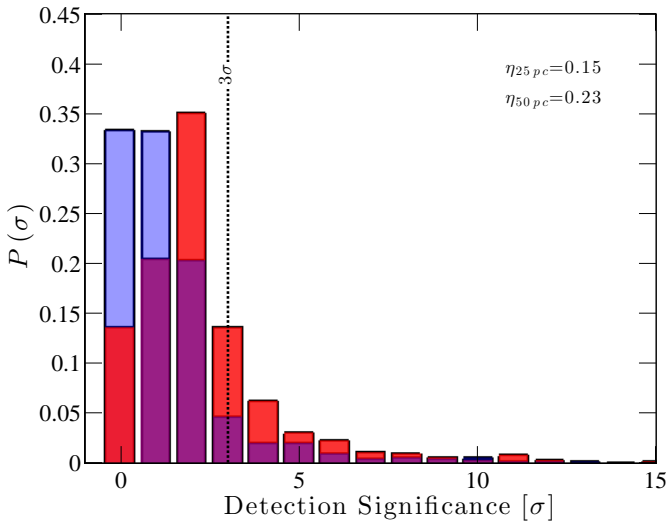}
\centering
\end{center}
\caption{Distribution of the detection significance of the central black hole masses in Monte Carlos simulations. The blue and red bars show the results for rings with radii of 25 pc and 50 pc respectively. 
\label{fig:histogram}}
\end{figure}

As can be seen in Figure \ref{fig:cross}, although the shape of detection cross sections vary for different lensing configurations, $\eta\sim0.1$ regardless of orientation and ellipticity. To generalize these results for systems with different parameters, we run Monte Carlo simulations by drawing values for lens redshift (normal distribution, mean=0.5, rms=0.2), source redshift (normal, mean=3, rms=1), lens ellipticity (flat distribution, $0\,$--$\,1$), orientation angle (flat, $0^{\circ}\, $--$\, 90^{\circ}$), and inclination angle (flat, $0^{\circ}\, $--$\, 90^{\circ}$). The source is uniformly placed inside the $\mu_f \geqslant3$ cross section. We perform these simulations for rings with radii of 25 and $50$ pc. Figure \ref{fig:histogram} shows  the distribution of detection significances for the 25 (blue) and $50$ pc (red) rings. Given our assumed intrinsic line flux $\sim15\, \%$ and $23\, \%$ of BH masses can be measured with a significance of $\geqslant3\sigma$.

\section{Discussion}
\label{sec:discuss}

In this work, the rotational velocities of gas at $r=50$ pc and $r=200$ pc are used to simultaneously fit for the depth of the potential of the galaxy and the mass of the BH.
If the slope of the density profile of the host galaxy is also treated as a free parameter, then more gas tracers at different radii are needed to simultaneously constrain the density slope, the depth of the potential, and the mass of the BH. 
\citet{Davis:13} used the stellar mass-to-light ratio as a free parameter to subtract the stellar contribution to the enclosed mass in the inner ring, which requires high resolution imaging at other wavebands. With high resolution, however, the error from not subtracting the stellar mass becomes small and thus gas observations alone may be accurate enough to establish the SMBH-galaxy relations. Here, we have focused on the possibility of resolving the rotational velocities at small radii, in the sphere of influence of the central BHs where the BH mass is comparable to the contribution of mass of other components (dark matter, stars, and interstellar medium).  We therefore chose a simple mass model for the source galaxy (isothermal profile) and fixed the slope of the density profile, noting that in real data, there is a large probability of observing many gas clumps at various radii, allowing a more detailed dynamical mass model with more degrees of freedom. We also note that a more complex gas morphology (e.g., clumpy or filamentary) should not impose any limitations to this method, since lens models can reconstruct the detailed morphology of gas in every spectral channel \citep[see e.g.][]{Riechers:08,hezaveh:13a}. 
Simulations \citep[e.g.][]{hopkins:12} and observations \citep[e.g.][]{tacconi:08}, however,  suggest that many such systems may not be dynamically cold, which could limit the application of this method. 

The Monte Carlo simulations indicate that the BH masses could be detected for rings with inclination angles $\gtrsim15^{\circ}$. Above this limit, the detection sensitivity does not strongly depend on the inclination angle.
We assumed an intrinsic line flux of $4\, \mathrm{mJy\, km\, s^{-1}}$. 
This is in agreement with the central fluxes obtained from CO observations of high redshift star forming galaxies \citep[e.g.][]{hodge:12}, if we assume an exponential disk for the gas and set the half-light radii to the observed values for lensed galaxies \citep{hezaveh:13b}. We note that the flux of the gas at these small radii is the largest source of uncertainty in our estimates.

Although the simulations in this work specifically addressed the capabilities of ALMA for measuring BH masses, they are also informative for other upcoming high-resolution instruments. In particular, thirty meter class telescopes (TMT, GMT, and E-ELT) with milli-arcsecond resolution (comparable to that of ALMA) and equipped with integral field units
 \citep[e.g. the Infrared Imaging Spectrometer, ][]{simard:13} should be capable of carrying out such measurements if the source galaxies contain bright ionized gas at small radii. Upcoming optical surveys are expected to find a wealth of galaxy-galaxy strong lenses \citep[e.g. $\gtrsim1500$ with the Large Synoptic Survey Telescope; ][Figure 12.6]{LSST:09}. With such large numbers, even a small yield fraction, $\eta$, can result in a statistically significant sample of measured BH masses.

When ALMA and thirty meter telescopes start operating at such high angular resolutions, an important issue to consider will be the selection of suitable targets for high-resolution follow-ups, by identifying sources whose central regions are highly magnified.
Assuming that the BHs reside at the center of the velocity field of the galaxies, this will likely be achievable by lens models of intermediate resolution observations of molecular and atomic lines. Such models can locate the position of the center of the velocity fields, allowing an estimation of the angular magnification at the central regions. Another approach could be lens modeling based on lower-resolution observations of emission that could be uniquely associated with the central regions (e.g., hard X-rays coming from AGNs, narrow, or broad lines, etc.). With only the positions and fluxes of lensed images of such emissions, lens models can estimate the angular magnifications.
If target selections are based on magnification, however, special care should be given to understanding the selection methods: selection based on magnification from a population with various source sizes (radii of bright circumnuclear molecular gas clumps) can result in strong selection biases against extended systems \citep{hezaveh:12a}. If a correlation between BH mass and the radial position of gas clumps exists, this can result in a bias in the BH mass statistics.

Finally, we mention that \citet{Davis:2014} report that resolving the velocity gradient at larger radii, beyond the sphere of influence of BHs, can still yield robust BH mass measurements. This implies that lensing magnification may allow even lower mass BHs to be accessible at high redshifts.

\section{Conclusion}
\label{sec:conclusion}
A small but non-negligible fraction of  strongly lensed high redshift galaxies have highly magnified central regions.
We find that with ALMA and thirty meter optical telescopes, with milli-arcsecond angular resolutions and spatially resolved spectroscopy capabilities, we will be able to measure rotational velocities of gas as close as 25 pc and 50 pc to the central BHs (in $\sim 15\%$ and $\sim20\%$ of systems respectively), comparable to the sphere of influence of BHs of mass $\gtrsim10^8 M_{\odot}$, allowing us to measure the mass of high redshift BHs using resolved gas dynamics. Given the expected large samples of submillimeter and optical galaxy-galaxy strong lens systems from millimeter and optical surveys, such systems may allow a significant measurement of the $M-\sigma$ relation at high redshifts, shedding light on the mechanisms behind this intriguing correlation.

\acknowledgements{
I am grateful to Ryan Keisler for carefully reading this manuscript.
I thank Gil Holder, Phil Marshall, Norm Murray, Roger Blandford, Neal Dalal, Tim Davis, Laurence Perreault Levasseur, John Carlstrom, Dan Marrone, and the SPT-SMG team for useful discussions and Calcul Quebec for providing me with computing resources.}

\end{document}